\documentclass{webofc}
\usepackage[varg]{txfonts}   
\begin{document}
\title{NLO electroweak radiative corrections for four-fermionic process at Belle~II}
%
%

\author{\firstname{Aleksandrs} \lastname{Aleksejevs}\inst{1}\fnsep\thanks{\email{aaleksejevs@grenfell.mun.ca
    }} \and
        \firstname{Svetlana} \lastname{Barkanova}\inst{2}\fnsep\thanks{\email{svetlana.barkanova@acadiau.ca
             }} \and
        \firstname{Vladimir} \lastname{Zykunov}\inst{3}\fnsep\thanks{\email{vladimir.zykunov@cern.ch
             }}
}

\institute{Memorial University of Newfoundland, Corner Brook, Canada 
\and
           Acadia University, Wolfville, Canada 
\and
           Belarusian State University of Transport, Gomel, Belarus;
           Joint Institute for Nuclear Research, Dubna, Russia
          }

\abstract{%
We discuss the next to the leading order (NLO) electroweak radiative corrections 
to the $e^- e^+ \rightarrow f^- f^+ (\gamma)$ cross section asymmetry, 
for polarized and unpolarized beam scenario. The left-right and forward-backward 
amplitudes, with and without radiative corrections, are evaluated and compared for 
various kinematics. The hard bremsstrahlung is included for arbitrary energy cuts. 
The radiative corrections are shown to be significant and having a non-trivial dependency 
on the kinematic conditions. The calculations are relevant for the ultra-precise low-energy 
experiment Belle~II planned at SuperKEKB.

}
\maketitle

\section{Introduction}
\label{intro}
The precision electroweak physics will be accessed at low energy by the upcoming experiments such as 
Belle-II at SuperKEKB ~\cite{BelleIITDR}, MOLLER at JLab \cite{MOLLER} and P2 at MESA \cite{P2} and 
play an important complimentary role to the direct new-physics search at the LHC. Belle II aims to 
determine the weak mixing angle at $\sqrt{s} =m(\Upsilon_{4S})=10.577$ GeV, with the same precision 
as the LEP/SLC measurements made at the $Z$-boson pole and for $e^- e^+ \rightarrow b \bar b$ but free 
from fragmentation uncertainties \cite{Bevan}. However, before the reliable information can be extracted 
from the experimental data, it is necessary to consider the higher-order effects, i.e. electroweak radiative 
corrections (EWC). The inclusion EWC is an indispensable part of any modern experiment, but will be of the 
most paramount importance for the upcoming ultra-precise measurements, and, in some cases, must include 
not only a full treatment of one-loop radiative corrections (NLO) but also the leading two-loop corrections (NNLO). 

The significant theoretical effort has been dedicated to NLO EWC to electron-positron annihilation for more than 
three decades now, starting from \cite{BH82} which discussed EWC for an arbitrary polarization. Later, 
the collaborations such as BHM and WOH \cite{hollik,BH84}, LEPTOP \cite{LEPTOP}, TOPAZ0 \cite{TOPAZ96} 
and ZFITTER \cite{ZF91,grup-bar2}, provided a new level of precision for EWC at the $Z$-pole required by 
the LEP and SLC colliders. More recent EWC of the post-LEP/SLC era are offered by KK \cite{KK} and 
SANC \cite{sanc-eeff} codes. Although quick and accessible, these generic ``ready-to-use'' codes may not 
be optimal for the new generation of precision experiments requiring a more custom approach. The previous 
work by our theory group \cite{ABIZ2010ABIKZ2013} dedicated to the parity-violating (PV) tests of the 
Standard Model showed that the exact analytical one-loop calculations using the computer algebra approach 
did not only increase the theoretical precision dramatically, but also gave us an opportunity to verify 
previous calculations done in other formalisms \cite{13-Au}. A more detailed analysis of the low- and 
high-energy asymptotic behavior of electroweak radiative corrections in polarized $e^- e^+ \to \mu^- \mu^+$ 
process can be found in \cite{ABBZ2016}.

The main goal of this paper is to calculate a full gauge-invariant set of
NLO EWC, both with computer algebra requiring no simplifications, using our semi-automatic approach (SAA) 
based on FeynArts, FormCalc and LoopTools \cite{Hahn}, and analytically (on paper), in a compact form, using 
the asymptotic methods \cite{hollik,13-Au}, and to compare the results making sure that our calculations are error-free. 
A complete set of NNLO EWC is not yet available and not included in the current analysis, but the work is already 
in progress by several groups (for example, \cite{UsNNLO2016} and \cite{RHill2016}) and planned to be completed 
in time for the upcoming experiments such as Belle II, MOLLER and P2.

\section{NLO electroweak corrections}

\begin{figure}[h]
\centering
\includegraphics[width=3.32cm,clip]{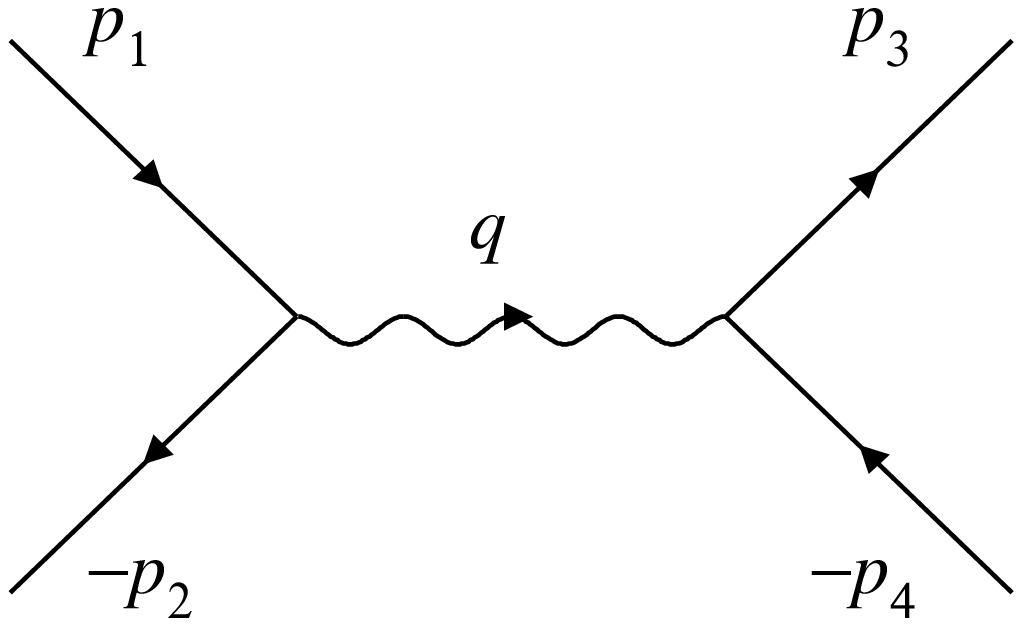}
\includegraphics[width=3.32cm,clip]{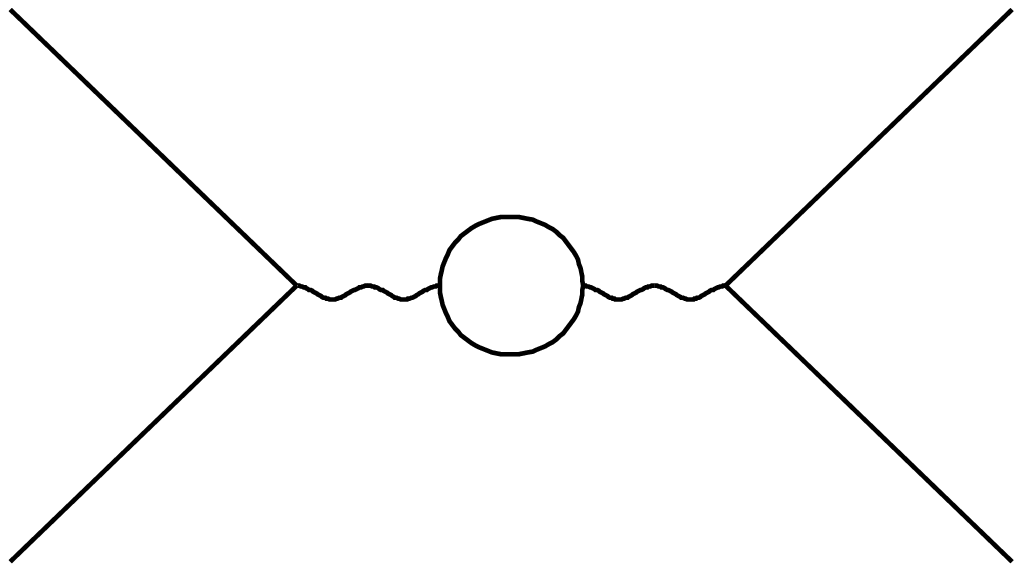}
\includegraphics[width=3.32cm,clip]{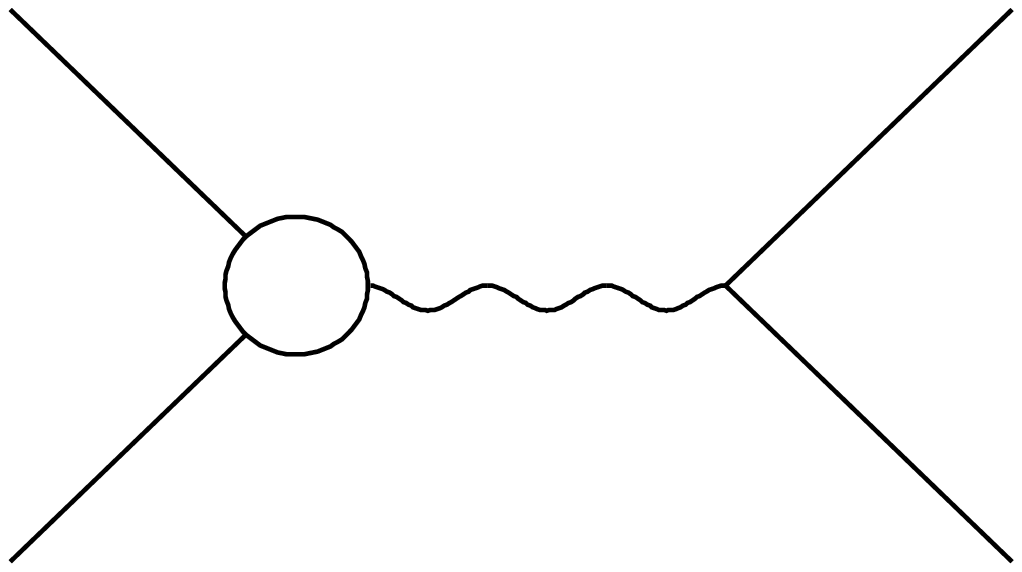}
\includegraphics[width=3.32cm,clip]{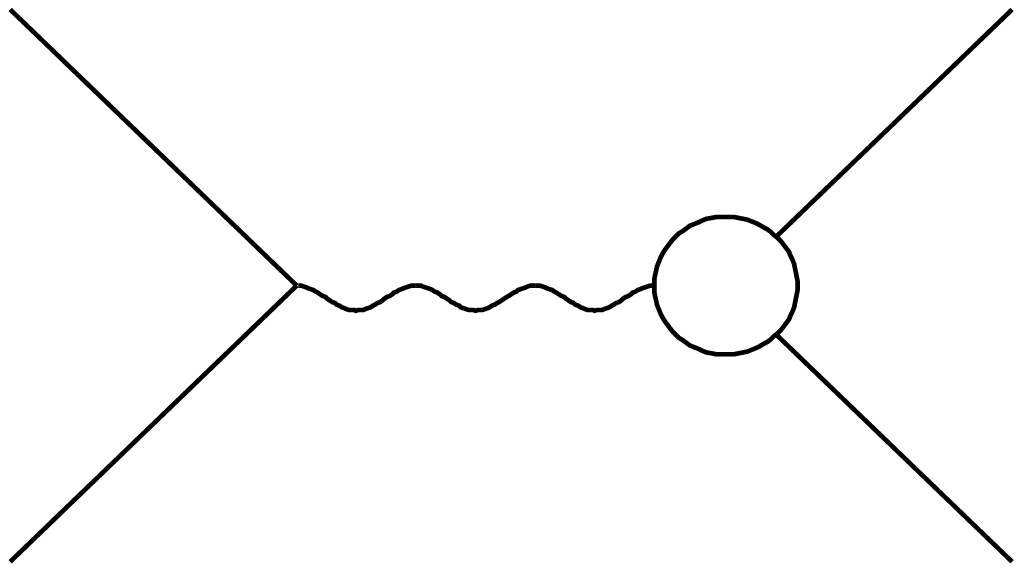}
\\[3mm]
\includegraphics[width=3.32cm,clip]{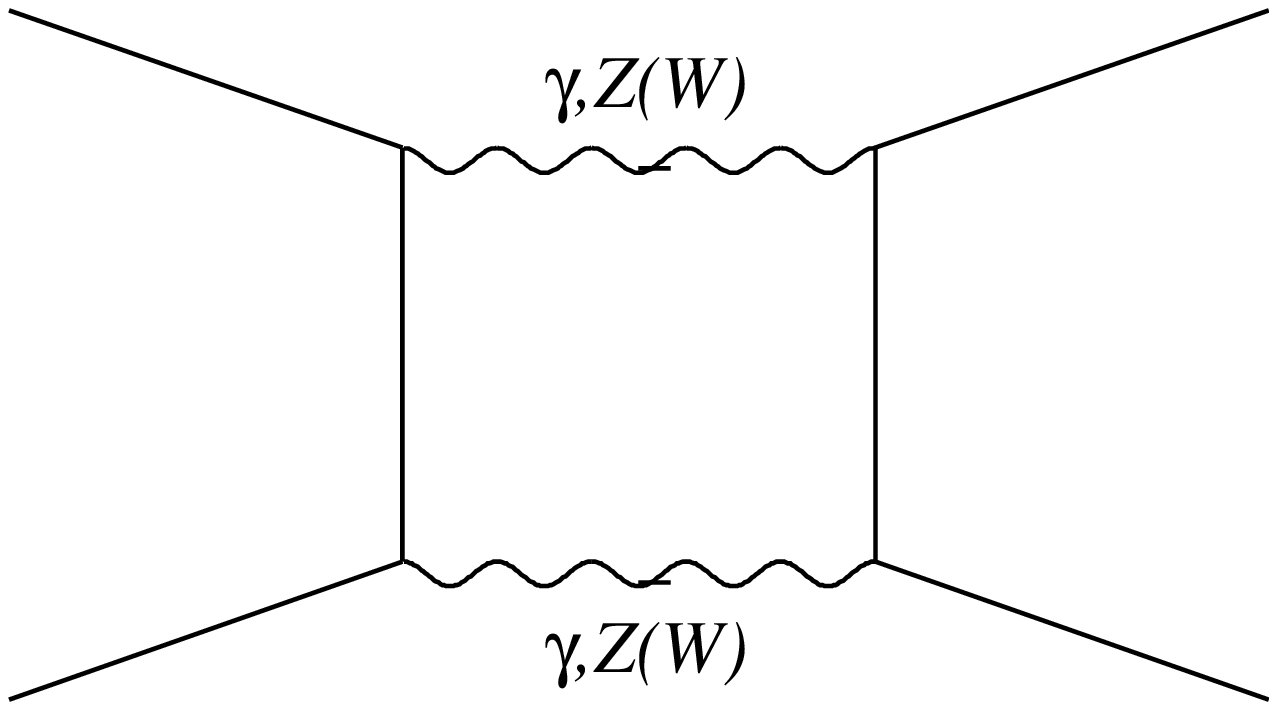}
\includegraphics[width=3.32cm,clip]{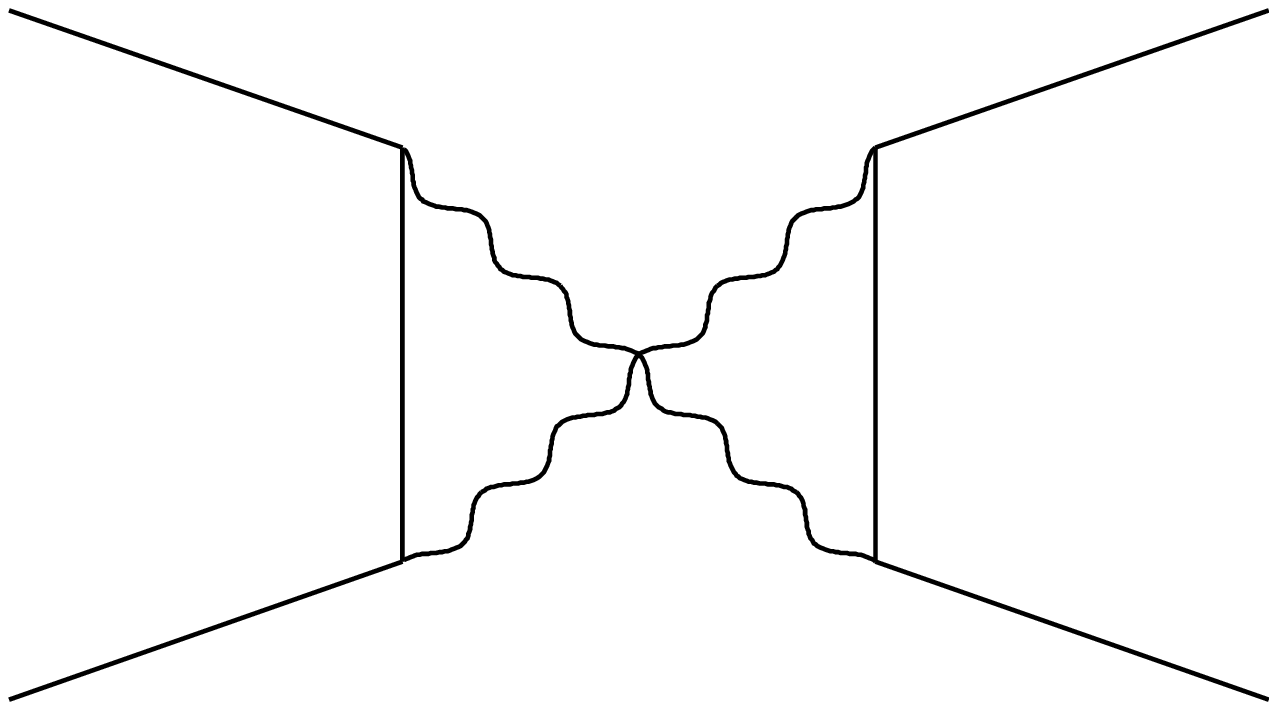}
\caption{
Feynman diagrams for the $e^- e^+ \rightarrow \mu^- \mu^+$ 
process in radiation-free kinematics
}
\label{born}       
\end{figure}

Let us start by defining the cross section for scattering of polarized electrons on unpolarized positrons,
\begin{equation}
e^-(p_1) + e^+(p_2) \rightarrow f^-(p_3)+f^+(p_4).
\label{0}
\end{equation}
In the Born approximation (leading order) illustrated by the first diagram in Fig.~\ref{born}, we have:
$\sigma^0 \approx \frac{\pi^3}{2s} |M_0|^2$. 
%
\smallskip
Here, $\sigma$ is the differential cross section $\sigma \equiv {d\sigma}/{d(\cos\theta)}$,
$\theta$  is the scattering angle of the detected muon
with 4-momentum $p_3$ in the center-of-mass system of the initial electron and positron, and $M_0$ is the Born (${\cal O}(\alpha)$) 
amplitude (matrix element). 
The 4-momenta of initial ($p_1$ and $p_2$) and final
($p_3$ and $p_4$) fermions generate a standard
set of Mandelstam variables:
\begin{equation}
s=(p_1+p_2)^2,\ t=(p_1-p_3)^2,\ u=(p_1-p_4)^2.
\label{stu}
\end{equation}

At the Next-to-Leading-Order (${\cal O}(\alpha^2)$), 
the differential cross section is given by the second term of the following expansion:
\begin{equation}
\sigma = \frac{\pi^3}{2s} |M_0+M_1|^2\approx \frac{\pi^3}{2s} (M_0M_0^+ + 2 {\rm Re} [M_1M_0^+] ),
\label{01a}
\end{equation}
where the one-loop amplitude $M_1$ is the sum of boson self-energy (BSE),
vertex (Ver) and box diagrams (see  Fig. \ref{born}):
\begin{equation}
M_{1}=M_{{\rm BSE}}+M_{{\rm Ver}}+M_{{\rm Box}}.
\end{equation}
There are no contributions from the electron self-energies in the on-shell renormalization scheme \cite{BSH86, Denner}.
The explicit form of the Born and one-loop amplitudes can be found in \cite{Yaf16}.

\begin{figure}[h]
\centering
\includegraphics[width=3.32cm,clip]{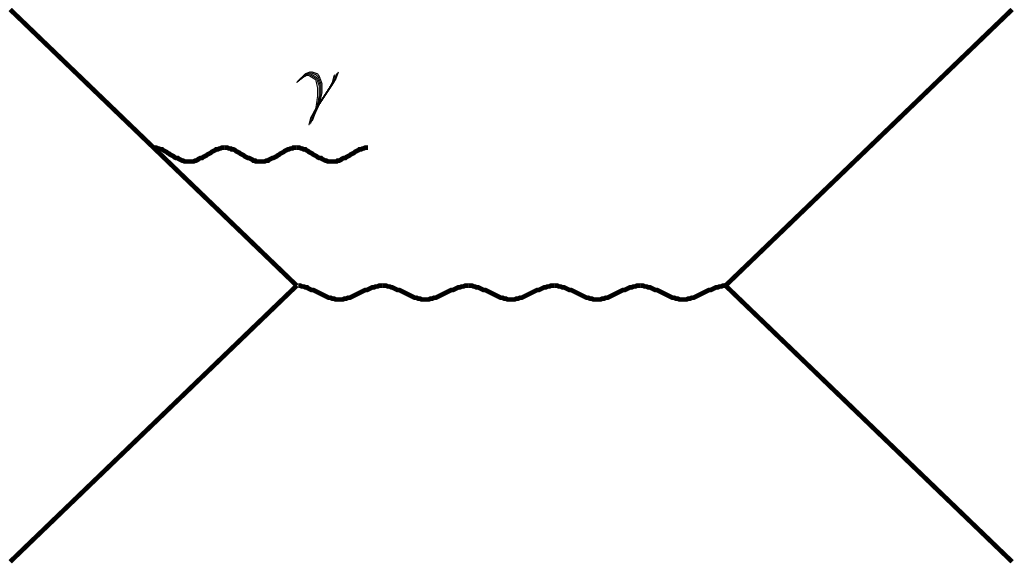}
\includegraphics[width=3.32cm,clip]{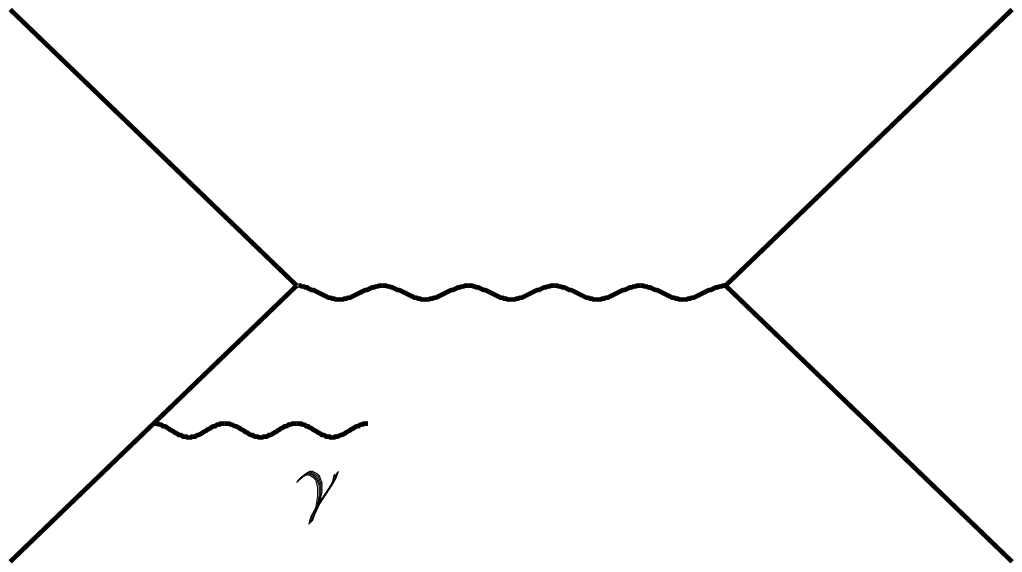}
\includegraphics[width=3.32cm,clip]{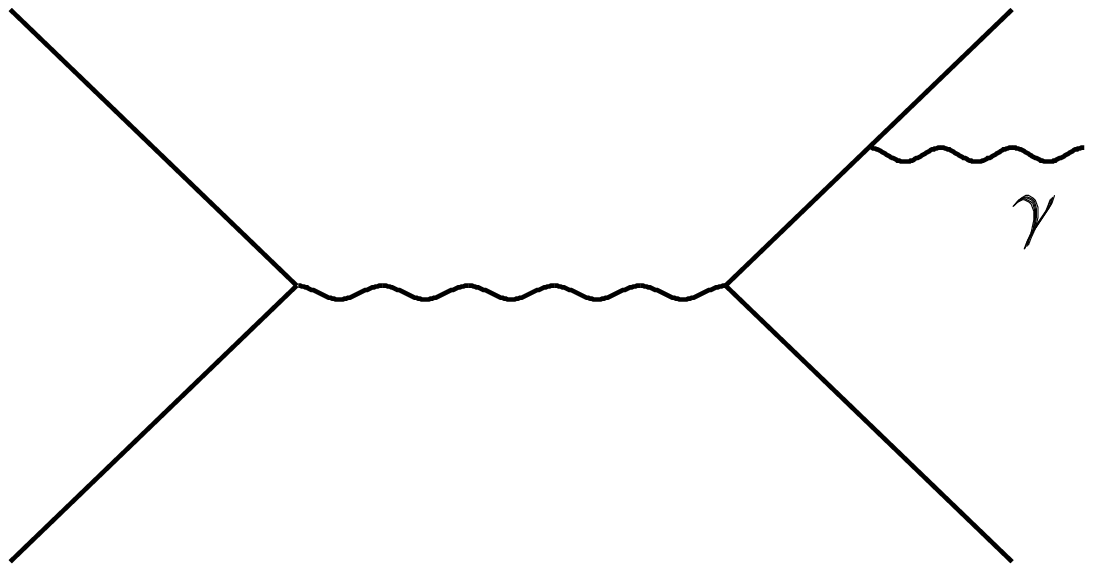}
\includegraphics[width=3.32cm,clip]{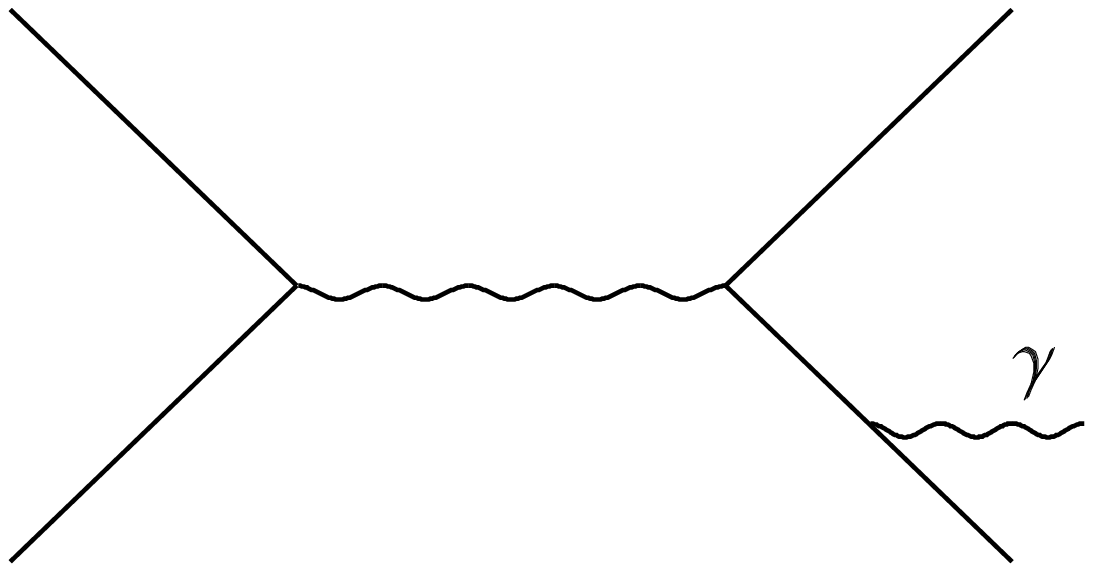}
\caption{
Diagrams with photon emission
}
\label{bre}       
\end{figure}

The bremsstrahlung diagrams corresponding to initial state radiation (ISR) and 
the final state radiation (FSR) are illustrated in Fig.~\ref{bre}.
\smallskip
The differential cross section for the process
\begin{equation}
e^-(p_1) + e^+(p_2) \rightarrow f^-(p_3)+f^+(p_4)+\gamma(p),
\label{BREM}
\end{equation}
has a form
\begin{equation}
d\sigma_R = \frac{\alpha^3}{\pi^2 s} \sum |R|^2 d\Gamma_3.
\label{sig-R}
\end{equation}
The amplitude $R$, the phase space of emitted photon $d\Gamma_3$, 
and explanation of infrared divergence are given in \cite{Yaf16}.

\section{Analysis}

For input parameters, we used $\alpha$,\ $m_W$,\  $m_Z$, $m_H$ and fermionic masses from \cite{PDG14}. 
The light quark masses reproducing $\Delta \alpha_{had}^{(5)}(m_Z^2)$=0.02757 \cite{jeger}, 
are regulated by the hadronic vacuum polarization, 
which does not introduce a significant uncertainty to our results. 
Please see \cite{ABIZ2010ABIKZ2013} for details.

\begin{figure}[h]
\centering
\includegraphics[width=5.94cm,clip]{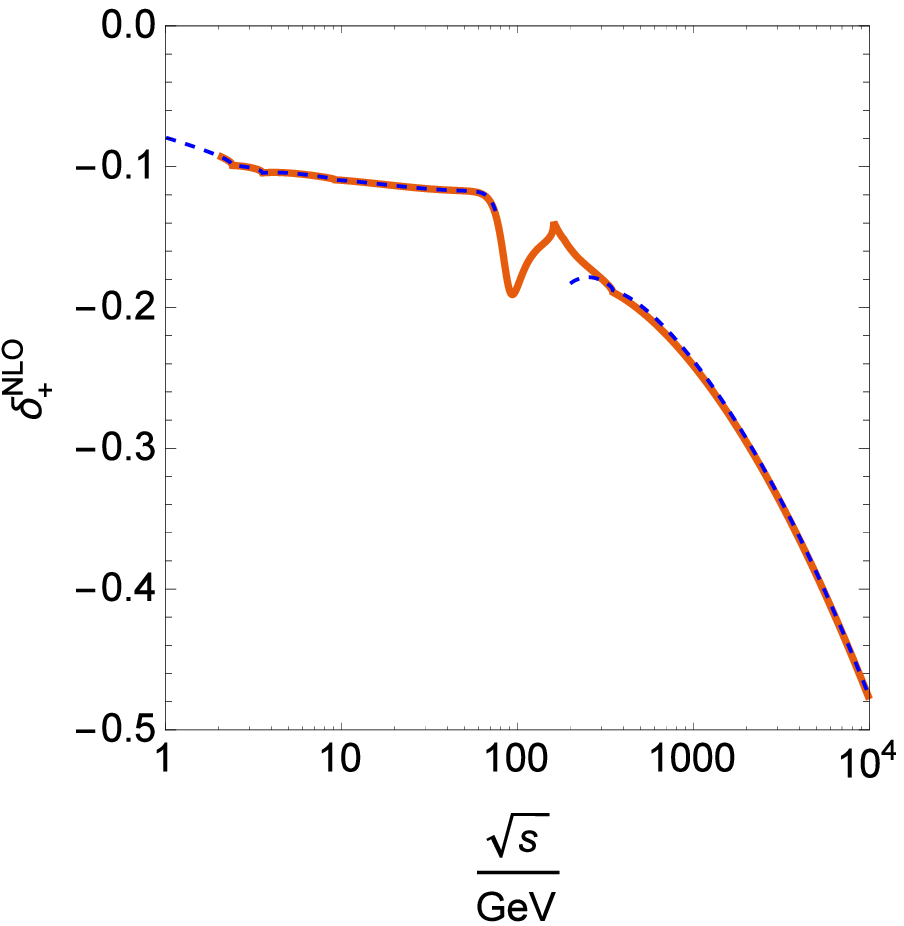}
\includegraphics[width=5.94cm,clip]{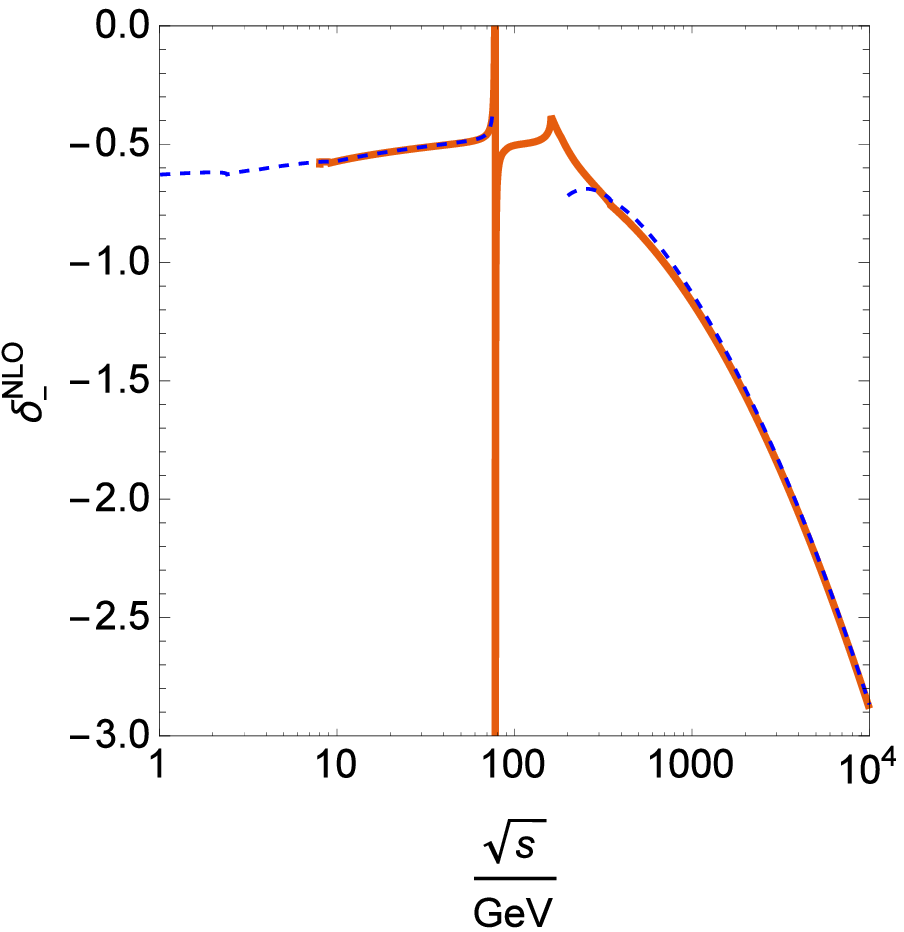}
\caption{
The relative NLO corrections at $\theta=90^\circ$ vs energy: 
solid line -- computer algebra code (not working at very small $\sqrt{s}$),
dashed line --  asymptotic estimations, good for $\sqrt{s}$ below $m_W$ 
and for $\sqrt{s}$ above $m_Z$.
}
\label{srav}       
\end{figure}

Let us start with a comparison between the asymptotic and full calculations,
using two relative corrections $\delta^{\rm NLO}_+ $ and  $\delta^{\rm NLO}_-$:
\begin{equation}
\delta^{\rm NLO}_{\pm} = 
 \frac{\sigma^{{\rm NLO}}_{L} \pm \sigma^{{\rm NLO}}_{R}}
      {\sigma^{0}_{L} \pm \sigma^{0}_{R}}.
\label{dpm}
\end{equation}            
They are additive, and allow to estimate the physical effect. The asymmetry can now be defined as:
\begin{equation}
\delta^{\rm NLO}_A=\frac{\delta^{\rm NLO}_- - \delta^{\rm NLO}_+}{1+\delta^{\rm NLO}_+}.
\label{dA}
\end{equation}            
The comparison between two methods is illustrated in Fig.~\ref{srav}.
For now, we apply the same cut on the soft photon emission energy as in \cite{5-DePo},
$\omega=0.05\sqrt{s}$, to be re-evaluated for each experiment specifically.


%
Let us denote the specific type of contribution in cross section or asymmetries by a subscript $C$. 
$C$ can be 0 (Born contribution), 1 (one-loop EWC contribution), or 0+1 (both), i.e.
$C=\{0, 1, 0\!\!+\!\!1\}$.
The PV (left-right) asymmetry,
the forward-backward asymmetry, and
the left-right asymmetry constructed from the integrated cross sections
are defined in the traditional way:
\begin{equation}
A^C_{LR} =
 \frac{\sigma^C_{L}-\sigma^C_{R}}
      {\sigma^C_{L}+\sigma^C_{R}},\ \ \
 A^C_{FB} =
  \frac{\Sigma^C_{F}-\Sigma^C_{B}}
       {\Sigma^C_{F}+\Sigma^C_{B}},\ \ \
 A^C_{LR\Sigma} =
  \frac{\Sigma^C_{L}-\Sigma^C_{R}}
       {\Sigma^C_{L}+\Sigma^C_{R}}.
 \label{AAA}
 \end{equation}
The subscripts $L$ and $R$ on the cross sections $\sigma \equiv {d\sigma}/{d(\cos\theta)}$ correspond
to the polarization degree of the electron $p_{B}$ = $-1$ and $p_{B}$ = $+1$ respectively.

The forward and backward cross sections defined as:
$$ \Sigma^C_{F} =\int\limits_0^{\cos a} \sigma^{C}_{00} \cdot d(\cos\theta),\ \ \
   \Sigma^C_{B} =\int\limits_{-\cos a}^0 \sigma^{C}_{00} \cdot d(\cos\theta).
$$
and left and right integrated (over $a \leq \theta \leq b$ segment) cross sections are:
$$ \Sigma^C_{L} =\int\limits_{\cos b}^{\cos a} \sigma^{C}_{L} \cdot d(\cos\theta),\ \ \
   \Sigma^C_{R} =\int\limits_{\cos b}^{\cos a} \sigma^{C}_{R} \cdot d(\cos\theta).
$$

\begin{figure}[h]
\centering
\includegraphics[width=5.94cm,clip]{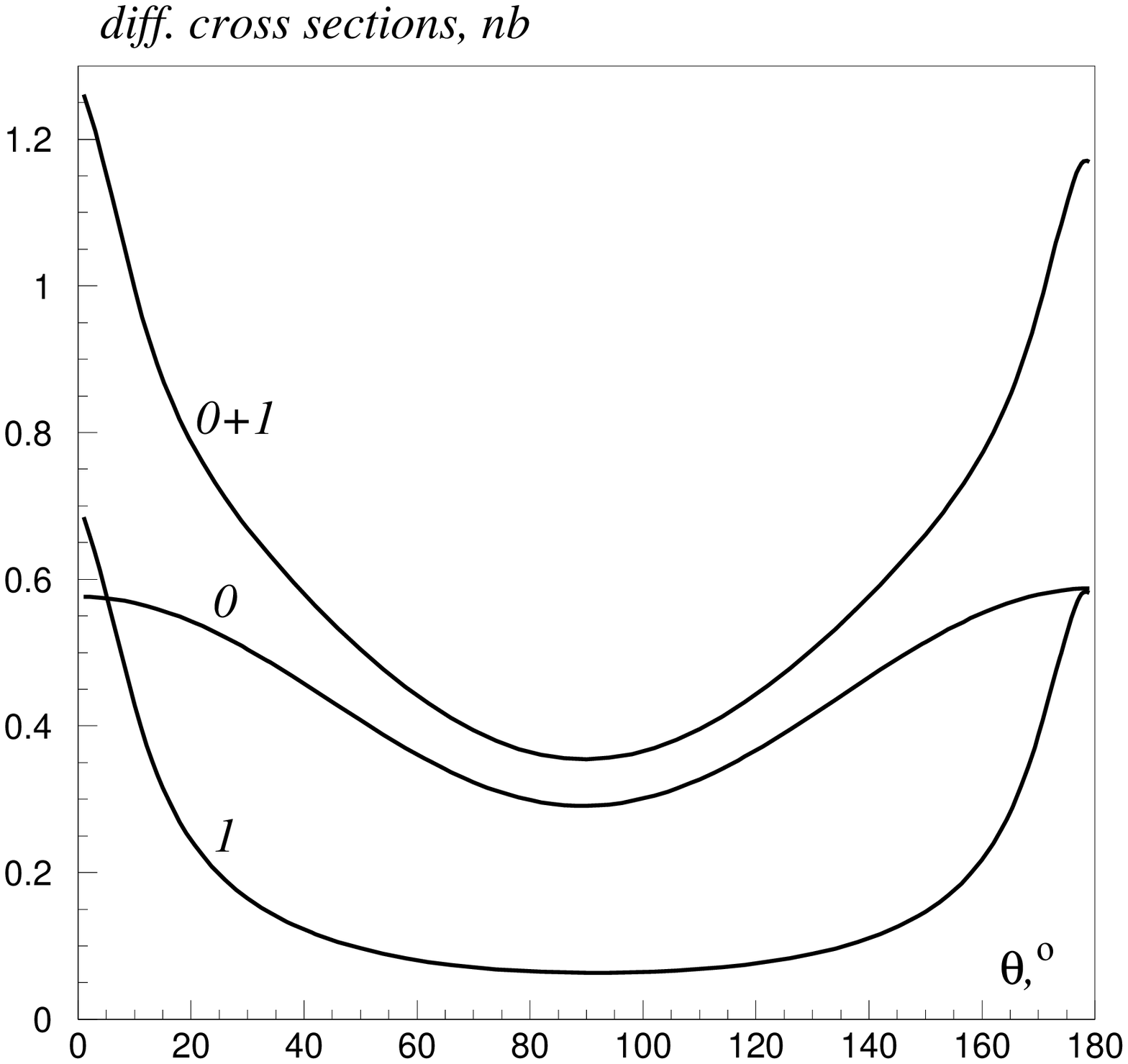}
\includegraphics[width=5.94cm,clip]{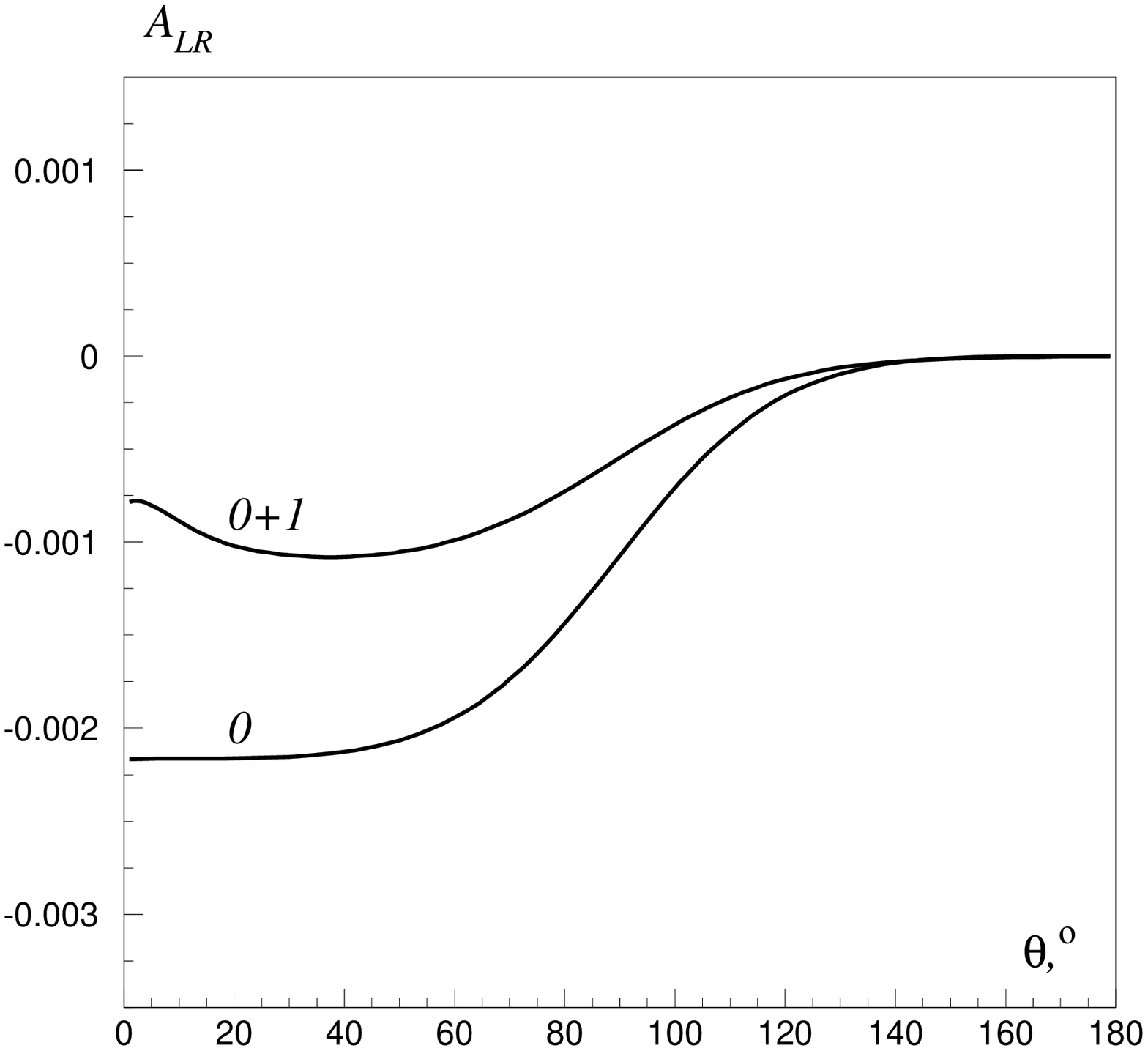}
\caption{
Left: unpolarized  
NLO corrected (0+1), Born (0), and their difference (1)
differential cross sections vs scattering angle $\theta$.
Right: the polarization Born asymmetry (0) 
and asymmetry taking into account the NLO EWC (0+1) vs scattering angle $\theta$.
Note the large contribution of EWC to $A_{LR}$ at small angles.
}
\label{a1}       
\end{figure}

\begin{figure}[h]
\centering
\includegraphics[width=5.94cm,clip]{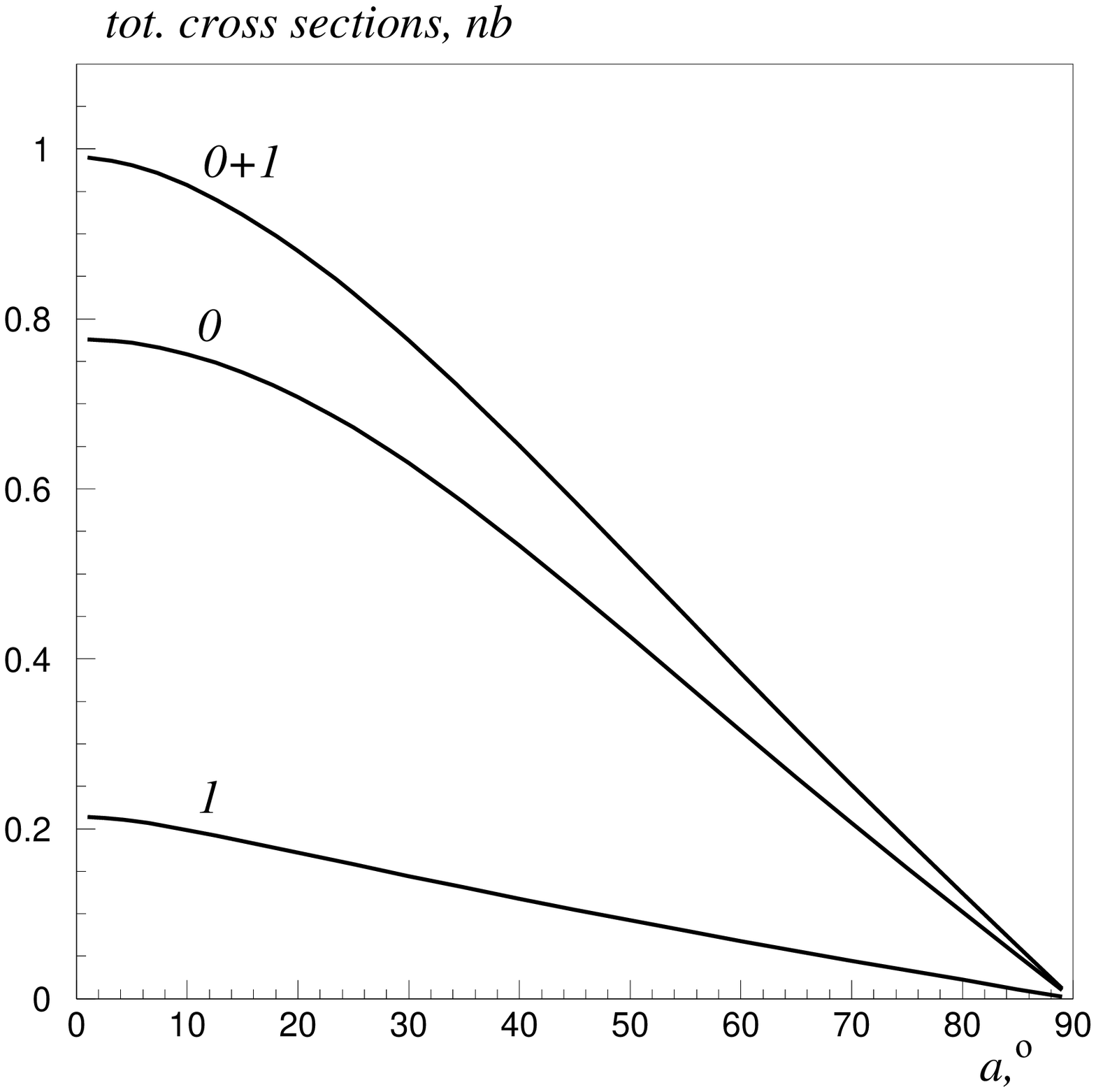}
\includegraphics[width=5.94cm,clip]{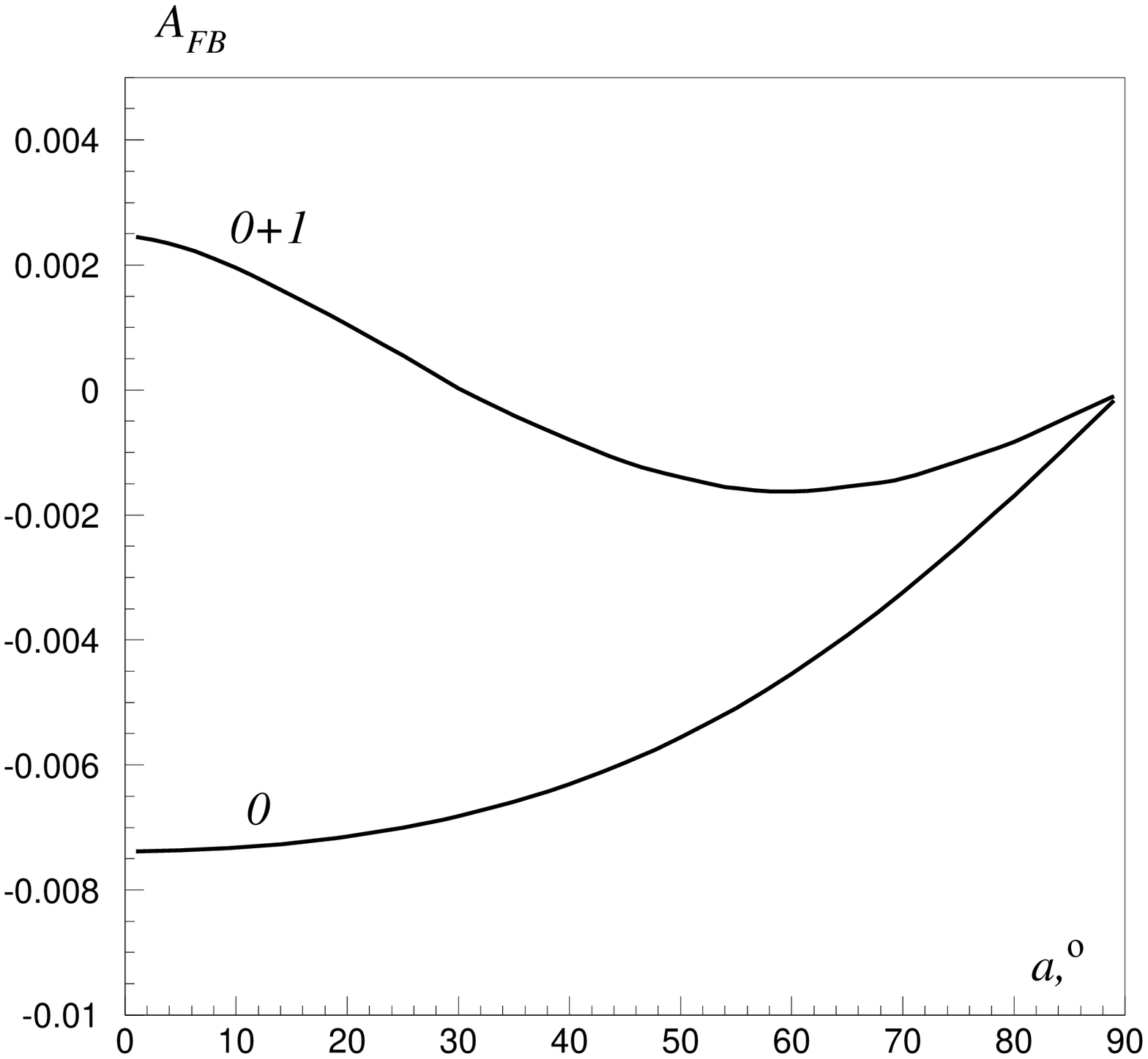}
\caption{
Left: unpolarized  NLO corrected (0+1), Born (0), and their difference (1) total cross sections vs angle $a$.
Right: the forward-backward Born asymmetry (0) 
and asymmetry taking into account the NLO EWC (0+1) vs angle $a$.
Note the importance of EWC.
}
\label{a2}       
\end{figure}

\begin{figure}[h]
\centering
\includegraphics[width=5.94cm,clip]{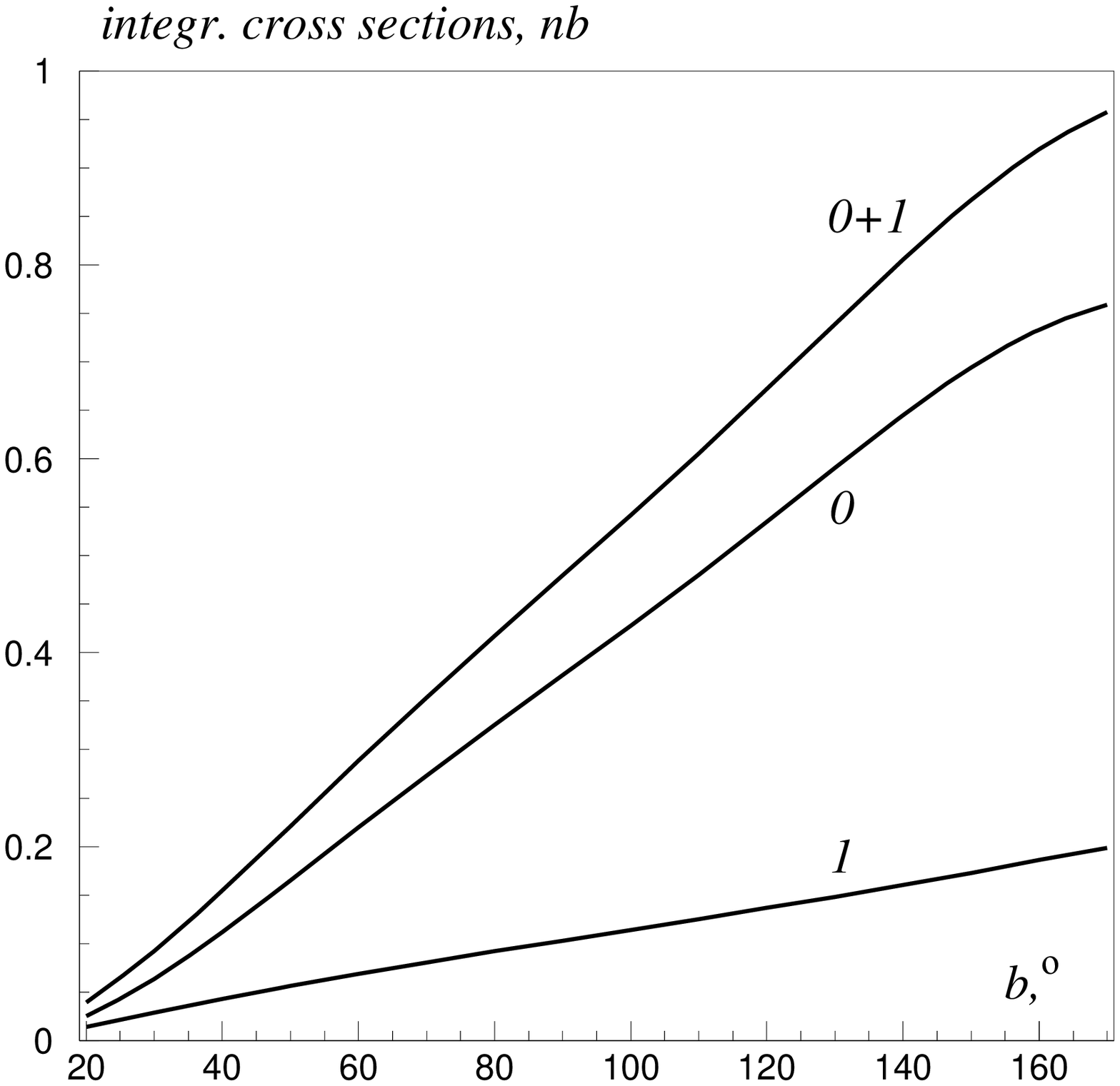}
\includegraphics[width=5.94cm,clip]{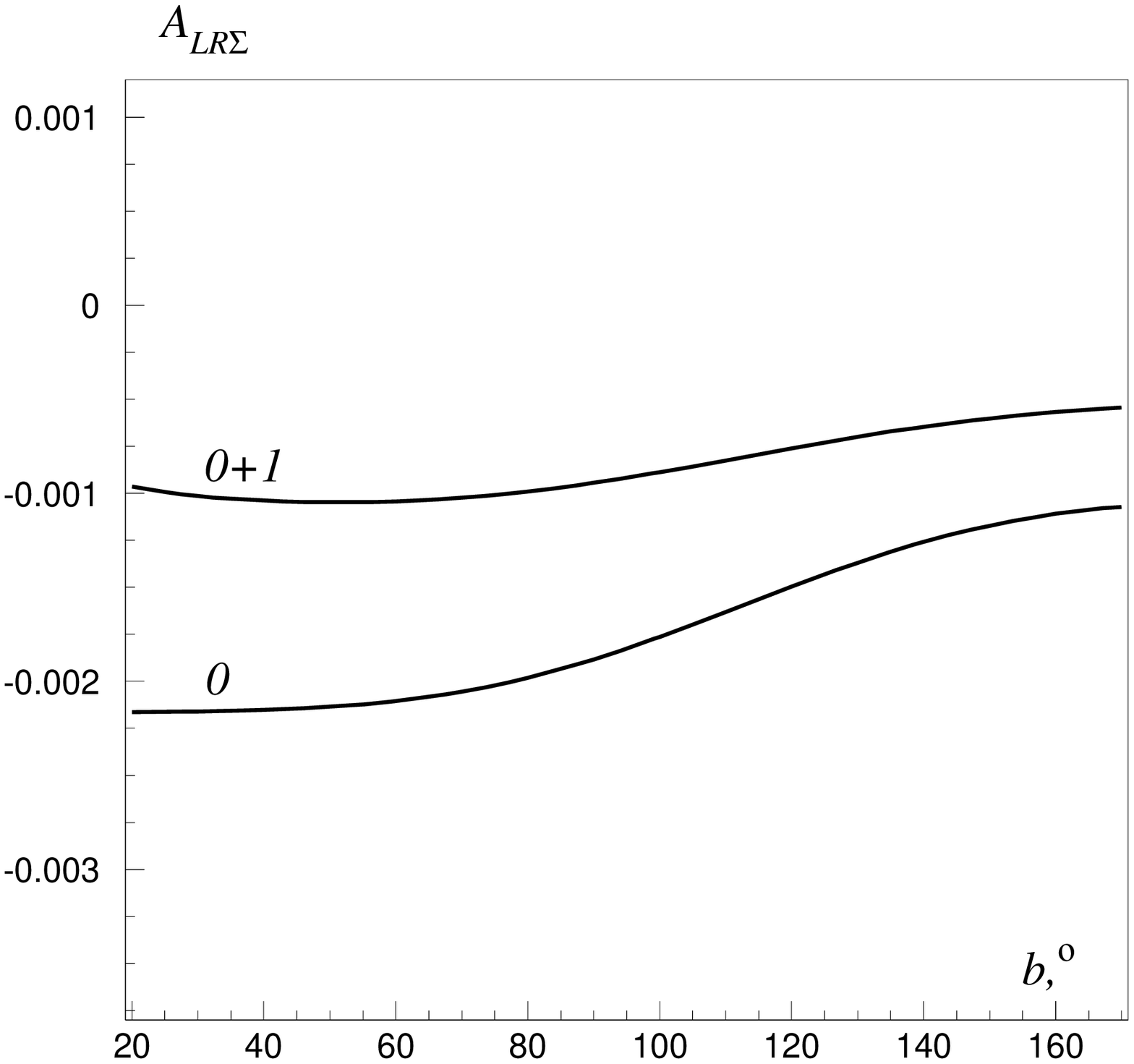}
\caption{
Left: unpolarized  
NLO corrected (0+1), 
Born (0),
and their difference (1) 
integrated cross sections vs angle $b$ at $a=10^\circ$.
Right: the left-right integrated Born asymmetry (0) 
and asymmetry taking into account the NLO EWC (0+1) vs angle $b$ at $a=10^\circ$.
}
\label{a3}       
\end{figure}

Figs.~\ref{a1}-\ref{a3} illustrate these asymmetries with and without the radiative effects.

\section{Conclusions}

We compare the results for the full set of one-loop EWC to left-right and forward-backward asymmetries at energies relevant for Belle II at SuperKEKB obtained by two different methods:
the exact (semi-automatic, with computer algebra) and approximate (asymptotic, on paper). 
Both methods have their advantages and limitations, but their combination proves that our work is error-free and provided more options to the experimental community. The bremsstrahlung process is fully controlled and the soft- and hard-photon approximations are compared. Although much improved in this work, the precision obtained with one-loop EWC is still insufficient for the upcoming experiments such as Belle II, MOLLER and P2 and will require at least the leading two-loop contributions.

\section{ACKNOWLEDGMENTS}
We are grateful to the organizers of Baldin ISHEPP XXIII Seminar for hospitality 
and to Yu. Bystritskiy, and J.M. Roney for stimulating discussions. 
V.Z. thanks Harrison McCain Foundation for support and the Acadia University for hospitality in 2016. 
The work of A.A. and S.B. is supported by the Natural Science and Engineering Research Council of Canada (NSERC).

\end{document}